\documentclass[12pt]{article}

\usepackage{epsfig}
\usepackage{graphicx}

\def\be{\begin{equation}}
\def\ee{\end{equation}}
\def\beq{\begin{eqnarray}}
\def\eeq{\end{eqnarray}}

\def\l{\lambda}

\def\r{\rho}

\def\vp{\varphi}

\def\N{{\cal N}}
\def\O{{\cal O}}

\def\({\left (}
\def\){\right )}
\def\[{\left [}
\def\[{\right ]}

\def\pr{{(\phi_{,r})}}

\begin{document}

\begin{titlepage}
\bigskip
\rightline{}
\rightline{hep-th/0409160}
\bigskip\bigskip\bigskip\bigskip
\centerline
{\Large \bf {An AdS Crunch in Supergravity}\footnote{To appear in
{\it The New Cosmology}, Proceedings of the Conference on Strings and 
Cosmology, Mitchell Institute, Texas A\&M University, March 14-17, 2004, 
eds R. Allen and C. Pope, to be published by the American Institute of 
Physics.}}
\bigskip\bigskip
\bigskip\bigskip
\centerline{\large Thomas 
Hertog\footnote{email:Hertog@vulcan.physics.ucsb.edu}}
\bigskip\bigskip
\centerline{\em Department of Physics, UCSB, Santa Barbara, CA 93106}
\bigskip\bigskip

\begin{abstract}

We review some properties of N=8 gauged supergravity in four dimensions with 
modified, but AdS invariant boundary conditions on the $m^2=-2$ scalars.
There is a one-parameter class of asymptotic conditions on these fields and the
metric components, for which the full AdS symmetry group is preserved. The 
generators of the asymptotic symmetries are finite, but acquire a contribution
from the scalar fields. For a large class of such boundary conditions, we find
there exist black holes with scalar hair that are specified by a single 
conserved charge. Since Schwarschild-AdS is a solution too for all boundary 
conditions, this provides an example of black hole non-uniqueness.
We also show there exist solutions where smooth initial data evolve to a big 
crunch singularity. This opens up the possibility of using the dual conformal 
field theory to obtain a fully quantum description of the cosmological 
singularity, and we report on a preliminary study of this. 

\end{abstract}

\end{titlepage}

\baselineskip=18pt

\setcounter{equation}{0}
%%%%%%%%%%%%%%%%%%%%%%%%%%%%%
\section{Introduction}
%%%%%%%%%%%%%%%%%%%%%%%%%%%%%

One of the main goals of quantum gravity is to provide a better understanding 
of the big bang or big crunch singularities in cosmology. An issue that 
immediately comes to mind is whether cosmological singularities represent a 
true beginning or end of evolution. If this is the case it would raise the 
question what determines the boundary conditions at the singularity. A truly 
unified theory should then, besides specifying the dynamics, also include a 
principle that specifies the universe's quantum state. An appealing proposal 
in this context is the quantum state given by the no boundary wave function 
\cite{Hartle83}. This describes the creation of an ensemble of universes with 
diverse properties. The no boundary proposal asserts that all information 
about a possible phase before the big bang that is in principle accessible to 
an observer in a given member of this ensemble is encoded in the no
boundary instanton. But there is no real sense in which evolution continues 
through the singularity: the instanton describes the beginning of a new, 
disconnected universe that has a self-contained physical description.  
Because there is no boundary in the past, the no boundary condition naturally 
leads to a top down approach to cosmology \cite{Hawking02}.
In this approach, one first specifies a number of properties (as few as 
necessary, of course) of the universe at late times, which are then used
to compute conditional probabilities predicting other features. The set of a 
posteriori conditons essentially select the histories that contribute to the 
path integral for a given member of the ensemble of universes.

Alternatively, it is possible that evolution continues through the 
singularity and that string theory itself determines the conditions at 
cosmological singularities. There may be some type of bounce, as envisioned 
by the pre-big bang \cite{Gasperini93} and cyclic universe models 
\cite{Khoury02}, or the transition could be chaotic, in which case one 
presumably needs to resort again to a top down approach to explain our
observed universe. Even if evolution continues through the singularity 
the quantum state at the singularity may contain certain universal 
features. Perhaps the correct answer will turn out to be a combination of both
scenarios: cosmological singularities could represent an endpoint of 
evolution only in certain situations, depending on the approach to the 
singularity. This would raise the interesting possibility that only certain 
`special' cosmologies could be created from a pre-big bang phase.

Since our usual notions of space and time are likely to
break down near cosmological singularities, a particularly 
promising approach to study this issue might be
to find a dual description in terms of more fundamental variables.
In string theory we do not yet have a dual description of real cosmologies, 
but we do have the celebrated AdS/CFT correspondence \cite{Maldacena98} which 
provides a non-perturbative definition of string theory on asymptotically 
anti-de Sitter (AdS) spacetimes in terms of a conformal field theory (CFT). 
The dual CFT description has been used to study the singularity inside black 
holes \cite{Fidkowski04}, which is  analogous to a cosmological singularity. 
Although some progress in this direction has been made,  the fact
that the singularity is hidden behind an event horizon clearly complicates 
the problem. This is because the CFT evolution is dual to bulk evolution in 
Schwarzschild time so the CFT never directly `sees' the singularity.

It would be better to have examples of solutions in a low energy supergravity 
limit of string theory where smooth, asymptotically AdS initial data evolve to
a big crunch singularity. Then AdS/CFT should provide a precise framework in
which the quantum nature of cosmological singularities could be understood,
at least with AdS boundary conditions. In this context, a big crunch 
singularity is simply any spacelike singularity which extends to infinity and 
reaches the boundary in finite time. 

In this lecture we present examples of such solutions in the abelian 
truncation 
of gauged $\N=8$, $D=4$ supergravity in which one focuses on the $U(1)^4$ 
Cartan subgroup of $SO(8)$. Gauged $N=8$ supergravity arises as the massless 
sector of eleven dimensional supergravity compactified on $S^7$. The 
truncation to the $U(1)^4$ sector contains three scalar fields with $m^2=-2$ 
in units of the AdS radius. We begin by reviewing the class of asymptotic 
conditions on these fields (and the metric components) that are invariant 
under the full AdS symmetry group. For each scalar we find (in addition to the
standard `Dirichlet' boundary conditions) there is a one parameter 
family of boundary conditions, labelled by $f$, that preserve the full set of 
AdS symmetries. When $f$ vanishes, the dual CFT is the usual $2+1$ theory
on a stack of M2-branes. Nonzero values of the parameter $f$ correspond to 
modifying this theory by a triple trace operator. On the bulk side, the 
generators of the asymptotic symmetries are finite for all $f$, but acquire a 
contribution from the scalar fields. 

For $f \neq 0$ we find there are static spherical black holes with scalar hair.
These solutions are specified by a single conserved charge, namely their mass.
Since Schwarschild-AdS is a solution too for all boundary conditions, this 
provides an example of black hole non-uniqueness. We then show there 
are also static solitons. We explain that the existence of solitons 
indicates AdS is nonlinearly unstable for these 
generalized AdS invariant boundary conditions. A particular manifestation of 
this is that for all nonzero $f$, there are bulk solutions where
smooth, finite mass initial data evolve to a big crunch. We conclude this 
lecture with
a preliminary discussion of the dual field theory description of the 
formation of a big crunch.

All this work was done in collaboration with G. Horowitz and K. Maeda, and the
reader is referred to the original papers for more details 
\cite{Hertog03c,Hertog04,Hertog04c}.
\setcounter{equation}{0}
%%%%%%%%%%%%%%%%%%%%%%%%%%%%%%%%%%%
\section{AdS Invariant Boundary Conditions}
%%%%%%%%%%%%%%%%%%%%%%%%%%%%%%%%%%

We first consider gravity in $d+1\ (d \geq 2)$ dimensions coupled to a single 
scalar 
field with a potential $V$ that has a negative maximum at $\phi=0$. This theory
admits a pure $AdS_{d+1}$ solution, with metric
\be \label{adsmetric}
ds^2_0 = \bar g_{\mu \nu} dx^{\mu} dx^{\nu}=
-(1+{r^2 \over l^2})dt^2 + {dr^2\over 1+r^2/l^2} + r^2 d\Omega_{d-1}
\ee
where the AdS radius is given by
\be
l^2= -\frac{d(d-1)}{2V(0)}
\ee
Since we are assuming that the scalar mass $m^2$ is less than zero, 
solutions to the linearized wave equation
$\nabla^2\phi -m^2\phi=0$ with harmonic time
dependence $e^{-i\omega t}$ all fall off asymptotically like 
\be\label{genfall}
\phi = {\alpha \over r^{\lambda_{-}}}  + {\beta \over r^{\lambda_{+}}}
\ee
with
\be\label{fallofftest}
\lambda_\pm = {d \pm \sqrt{d^2 + 4l^2 m^2}\over 2}
\ee
where we are assuming $m^2 \geq  -{d^2\over 4l^2} = m^2_{BF}$.
For fields that saturate the Breitenlohner-Freedman (BF) bound
\cite{Breitenlohner82}, $\lambda_+ = \lambda_- \equiv \lambda$ 
and the second solution asymptotically behaves like $\ln r/r^{\lambda}$.

We are interested in this lecture in nonlinear 
perturbations of (\ref{adsmetric})
where the scalar asymptotically behaves as (\ref{genfall}). Asymptotically 
anti-de Sitter spacetimes are defined by a set of boundary
conditions at spacelike infinity which satisfy the requirements set out in 
\cite{Henneaux85}. The standard set of boundary conditions  on the metric 
components \cite{Henneaux85} that are left invariant under $SO(d-1,2)$ are
\beq \label{usual}
g_{rr} = {l^2 \over r^2} -{l^4 \over r^4} +O(1/r^{d+1})  
& \qquad \ \ \ \ \ \ \ \ \ \ \ \ 
g_{tt}= -{r^2 \over l^2} -1 +O(1/r^{d-3})\nonumber\\
g_{tr}=O(1/r^{d})\ \ \  & \qquad g_{ra}=O(1/r^{d})\ \  \ \ \ \nonumber\\
g_{at}=O(1/r^{d-3}) & \qquad \ \ \ \ \ g_{ab}=\bar g_{ab}+O(1/r^{d-3})
\eeq
These boundary conditions go together with (and indeed require)
the standard `Dirichlet' boundary conditions on the scalar field, which
amount to taking $\alpha=0$ in (\ref{genfall}). It is well known that with 
these boundary conditions, a scalar field with negative mass 
squared does not cause an instability in anti de Sitter space. 
For boundary conditions of this form
there is a positive energy theorem \cite{Abbott82,Gibbons83,Townsend84} which 
ensures that the total energy cannot be negative as long as the scalar
does not violate the BF bound. 

Recall that the energy, and more generally, conserved charges associated
with asymptotic symmetries $\xi^\mu$ can be defined as follows 
\cite{Henneaux85}.  One starts with the Hamiltonian (we have set $8\pi G=1$)
\be
H[\xi] = \int d^dx \xi^\mu H_\mu= 
\int d^{d}x(\xi^{\perp}{\cal H}_\perp(x)+\xi^i {\cal H}_i(x))
\ee
where $H_\mu$ are the usual Hamiltonian and momentum constraints, 
\beq 
{\cal H}_\perp & = &\frac{2}{\sqrt{g}}(\pi^{ij}\pi_{ij}-
\frac{\pi^2}{d-1}+\frac{p^2}{4})
+\sqrt{g}\left[-\frac{R}{2}+\frac{1}{2}(D\phi)^2+V(\phi)\right],
\nonumber \\
{\cal H}_i & = & -2\sqrt{g}D_j\left(\frac{{\pi^j}_i}{\sqrt{g}}\right)
+pD_i\phi
\eeq
and $\pi^{ij}$ and $p$ are the momenta conjugate to 
$g_{ij}$ and $\phi$.
One then adds surface terms so that $H$ has well defined
functional derivatives, and one subtracts the analogous expression for
the $AdS_{d+1}$ background. 
For $\alpha=0$ boundary conditions on the scalar field (together with 
(\ref{usual})), this procedure yields the standard `gravitational' surface term,
\be \label{gravcharge}
Q_G[\xi]= \frac{1}{2}\oint dS_i
\bar G^{ijkl}(\xi^\perp \bar{D}_j h_{kl}-h_{kl}\bar{D}_j\xi^\perp)
+2\oint dS_i\frac{\xi^j {\pi^i}_j}{\sqrt{\bar g}}
\ee
where $G^{ijkl}={1 \over 2} g^{1/2} (g^{ik}g^{jl}+g^{il}g^{jk}-2g^{ij}g^{kl})$,
$h_{ij}=g_{ij}-\bar{g}_{ij}$ is the deviation from the spatial metric 
$\bar{g}_{ij}$ of pure AdS, $\bar{D}_i$ denotes covariant differentiation 
with respect to $\bar{g}_{ij}$ and $\xi^\perp = \xi^{\mu} n_{\mu}$ with
$n_{\mu}$ the unit normal to the surface. 

However, for scalar fields with $m^2$ in the range 
$ m^2_{BF} +1 >m^2 >m^2_{BF}$ we have
recently found there exists an additional one-parameter family of AdS 
invariant boundary
conditions on the scalar field and the metric components \cite{Hertog04}. 
More precisely, we find that 
the asymptotic AdS symmetries are also preserved in
solutions that belong to the following class,
\be\label{falloffphi}
\phi (r,t,x^a) = {\alpha (t,x^a) \over r^{\lambda_{-}}} +
{f \alpha^{\lambda_{+}/\lambda_{-}} (t,x^a) \over r^{\lambda_{+}}}
\ee
\beq \label{falloffg}
g_{rr}=\frac{l^2}{r^2}-\frac{l^4}{r^4}
-{\alpha^2 l^2\lambda_{-} \over (d-1)r^{2+2\lambda_{-}}}
+O(1/r^{d+2})\ \  
& \quad g_{tt}=-{ r^2 \over l^2} -1+O(1/r^{d-2}) \nonumber\\
g_{tr}=O(1/r^{d-1}) \qquad \qquad \qquad & \ 
g_{ab}= \bar g_{ab} +O(1/r^{d-2}) 
\nonumber\\
g_{ra} = O(1/r^{d-1}) \qquad \qquad \qquad & g_{ta}=O(1/r^{d-2}) \ \ \ \ \ \  
\eeq
where $x^a$ labels the coordinates on $S^{d-1}$ and 
$f$ is an arbitrary constant 
that labels the different boundary conditions. Notice that the
boundary conditions on some of the metric components are relaxed compared
to the standard set. For $f=0$ we recover boundary conditions on the
scalar corresponding to $\beta =0$ in (\ref{genfall}),
which have been considered previously 
in the context of AdS/CFT \cite{Klebanov99}. Remarkably, 
however, the full AdS symmetry group is preserved for all 
values of $f$. In particular, it is easy to see that
rescaling $r$ leaves $f$ unchanged. Since $\alpha$ depends on the
particular solution and can vanish, each of these boundary conditions admits
$AdS_{d+1}$ as a solution.

For these more general boundary conditions, the usual energy 
(\ref{gravcharge}) diverges as $r^{d-2\lambda_{-}}$. However, the purely
gravitational surface 
term (\ref{gravcharge}) no longer equals the conserved charge associated with the 
asymptotic symmetry $\xi=\partial_{t}$. Instead, by repeating the above 
procedure, one finds the conserved charges acquire an additional 
contribution from the scalar field.
The conserved charges now read \cite{Hertog04}
\be
\label{totcharge}
Q[\xi]=Q_G[\xi]+{1 \over 2d} \oint \xi^\perp 
\left[ (\nabla \phi)^2 -m^2\phi^2 \right].
\ee
For all finite $f$ (including $f=0$!) the scalar and gravitational terms 
separately diverge. The divergences, however, exactly cancel out yielding
finite total charges $Q[\xi]$. By contrast, the scalar charges $Q_\phi$ 
vanish for the standard $\alpha=0$ scalar boundary conditions.
For the case $f=0$ the scalar surface term is equivalent to the surface term 
$-{1 \over 2} \oint \phi \nabla_i \phi dS^i $ introduced by 
Klebanov and Witten in $D=4$ supergravity, to regularize the action of the 
$\alpha/r$ modes of the $m^2=-2$ scalar  \cite{Klebanov99}.

For spherical solutions that are asymptotically of the form
(\ref{falloffphi})-(\ref{falloffg}), it is easy to compute the total mass 
$M$. One obtains 
\be \label{totmassspher}
M=Q[\partial_t]=  \mathrm{Vol}(S^{d-1})\left( {d-1 \over 2} M_0 
- {2f m^2 \alpha^{d/\Delta_{-}} \over d} \right),
\ee
where $M_0$ is the coefficient of the $O(1/r^{d+2})$ correction to the
$g_{rr}$ component of the AdS metric.
We emphasize again that in the theory defined by $f=0$ boundary conditions, 
which is 
often used in AdS/CFT, the backreaction of the scalar relaxes 
the asymptotic falloff of some metric components, while
preserving the asymptotic 
AdS symmetry group. Although there is no residual finite
scalar contribution to the total mass $M$ in this case, it is only the
variation of the sum of both charges that is well defined.

Finally we briefly mention the case of a scalar saturating the BF 
bound, which generically behaves as $\ln r/r^{\lambda}$ near the boundary.
One finds there is again a one-parameter family of boundary conditions, 
involving the logarithmic branch, that preserves the AdS symmetries
\cite{Hertog04,Henneaux04}.
For all finite values of the parameter $f$ that labels the different
asymptotic conditions, the gravitational and scalar 
surface terms are logarithmically divergent. The divergences again cancel 
out, however, rendering the total charges (\ref{totcharge}) finite. 

\setcounter{equation}{0}
%%%%%%%%%%%%%%%%%%%%%%
\section{$D=4$ Gauged Supergravity}
%%%%%%%%%%%%%%%%%%%%%%%

We now consider the low energy limit of string theory
with $AdS_4 \times S^7$ boundary conditions. The massless sector 
of the compactification of $D=11$ supergravity on $S^7$ is 
${\cal N}=8$ gauged supergravity in four dimensions \cite{deWit82}.
The bosonic part of this theory
involves the graviton, 28 gauge bosons in the adjoint of $SO(8)$,
70 real scalars, and it admits $AdS_4$ as a vacuum solution.
It is possible to consistently truncate this theory to its
abelian $U(1)^4$ sector \cite{Duff99}. The resulting action is given by
\be\label{4-action}
S=\int d^4x\sqrt{-g}\left(\frac{1}{2}R
-\frac{1}{2}\sum_{i=1}^3 [(\nabla\phi_{i})^2 -2\cosh(\sqrt{2}\phi_{i})]\right)
+...
\ee
where the dots refer to gauge field terms that will be set to zero in this 
paper.
Here we have chosen the gauge coupling so that the AdS radius is equal to one.
Notice that the potential is unbounded from below, and the scalars have mass
\be
m^2_{i} = -2 \ .
\ee
The BF bound in four dimensions is $m_{BF}^2 = -9/4$.
Therefore, in addition to the standard Dirichlet boundary conditions where
asymptotically $\phi_{i} \sim \beta_{i}/r^2$, there is a class of 
asymptotically AdS solutions of the form
\be 
\label{4-scalar}
\phi_{i}(r,t,x^{a})=\frac{\alpha_{i} (t,x^a)}{r}+
\frac{f \alpha^2_{i} (t,x^a)}{r^2}
\ee
and
\beq 
\label{4-grr}
g_{rr}=\frac{1}{r^2}-\sum_{i=1}^3 \frac{(1+\alpha^2_{i}/2)}{r^4}+
O(1/r^5) & \quad g_{tt}=-r^2 -1+O(1/r) \nonumber\\
g_{tr}=O(1/r^2) \qquad \qquad \qquad & \ \  \ \ \ g_{ab}= \bar g_{ab} +O(1/r) 
\nonumber\\
g_{ra} = O(1/r^2) \qquad \qquad \qquad & g_{ta}=O(1/r) \ \ 
\eeq
where $x^a=\theta,\phi$ and $f$ is an arbitrary constant labelling the
different theories. The conserved charges for these boundary 
conditions acquire a scalar contribution and take the form
\be
\label{charge4d}
Q[\xi]=Q_G[\xi]+{1 \over 6} \sum_{i=1}^3 \oint \xi^\perp 
\left[ (\nabla \phi_{i})^2 +2\phi^2_{i} \right].    
\ee
We now turn to a more detailed analysis of this theory,
with boundary conditions specified by (\ref{4-scalar})-(\ref{4-grr}). 
To simplify the analysis we concentrate on solutions with only
one nontrivial scalar $\phi$.
\setcounter{equation}{0}
%%%%%%%%%%%%%%%%%%%
\section{Black Holes with Scalar Hair}
%%%%%%%%%%%%%%%%%%%

First we look for static, spherically symmetric AdS black hole solutions with 
scalar hair.
The original no hair theorem of Bekenstein \cite{Bekenstein74} proves there
are no asymptotically flat black hole solutions with scalar hair for minimally 
coupled scalar fields with convex potentials. This result was generalized 
to the 
case of minimally coupled scalar fields with arbitrary positive potentials in 
\cite{Heusler92}. Later it was shown 
\cite{Sudarsky02} there are no hairy, asymptotically AdS black holes where 
the scalar field asymptotically tends to the true minimum of the potential. In 
\cite{Torii:2001pg}, however, an example was given of a hairy black hole where 
the scalar field asymptotically goes to a negative maximum of the potential.
It is, however, not clear this solution can be regarded as being 
asymptotically AdS in a meaningful way, because its mass diverges. 

More recently, however,
a one-parameter family of AdS black holes with scalar hair was
found in three dimensions \cite{Henneaux02}. Asymptotically the scalar field 
again tends to a negative maximum, but the potential satisfies the 
BF bound and the solutions belong to the 
class (\ref{falloffphi})-(\ref{falloffg}) in three dimensions.
This raises the question if AdS black holes with scalar hair also 
exist in supergravity in four dimensions. 
In particular, it is possible that Bekenstein's no hair theorem applies to
supergravity with some AdS invariant boundary conditions, but not with others.

Writing the metric as
\be
ds_4^2=-h(r)e^{-2\delta(r)}dt^2+h^{-1}(r)dr^2+r^2d\Omega_2^2
\ee
the field equations read
\be\label{bhhairy14d}
h\phi_{,rr}+\left(\frac{2h}{r}+\frac{r}{2}\phi_{,r}^2h+h_{,r}
\right)\phi_{,r}   =  V_{,\phi}
\ee
\be\label{bhhairy24d}
1-h-rh_{,r}-\frac{r^2}{2}\phi_{,r}^2h =  r^2V(\phi)
\ee
\be\label{bhhairy34d}
\delta_{,r} = -{ r \phi_{,r}^2 \over 2}
\ee

%%%%%%%%%%%%%%%%%%%%%%%%%%%%%%%
\begin{figure}[htb]
\begin{picture}(0,0)
\put(56,254){$\phi_e$}
\put(412,23){$R_{e}$}
\end{picture}
%\centering{\psfig{file=bh4dhair.eps,width=5.5in}}
\mbox{\epsfxsize=14cm \epsfysize=9cm \epsffile{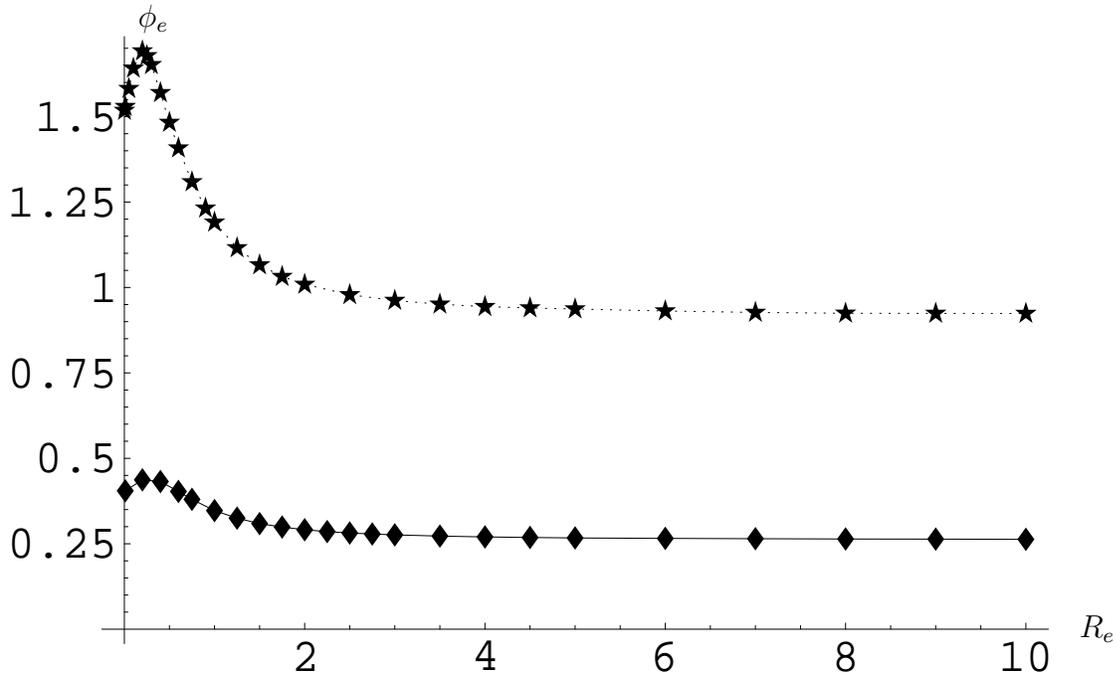}}
\caption{The value of the scalar field, $\phi_e$, at the horizon of a hairy 
black hole as a function of horizon size $R_e$.
The two curves show two one-parameter family of solutions of 
$D=4$ ${\cal N}=8$ supergravity, with two different AdS invariant 
boundary conditions, namely $f=-1$ (bottom) and $f=-1/4$ (top).}
\label{1}
\end{figure}
%%%%%%%%%%%%%%%%%%%%%%%%%%%%%%%%%%%

Regularity at the event horizon $R_e$ requires
\be \label{horcon4d}
\phi'(R_{e}) = {R_{e}V_{,\phi}\over 1-R_{e}^2V}
\ee
Asymptotic AdS invariance requires $\phi$ asymptotically decays as
\be \label{hair4d}
\phi(r)=\frac{\alpha}{r}+\frac{f\alpha^2}{r^2}, 
\ee
where $f$ is a given constant that is determined by the choice of
boundary conditions. Hence asymptotically
\be \label{asmetric4d}
h(r)=r^2+1+\alpha^2/2 -\frac{M_0}{r}, 
\ee
where $M_0$ is an integration constant.

The Schwarschild-AdS black hole with $\phi=0$ everywhere 
outside the horizon is a solution for all AdS invariant 
boundary conditions. Its mass (\ref{charge4d}) is given by
\be\label{mass4dschw}
M_s=Q[\partial_{t}]=4\pi M_0 =4\pi (R_e^3 +R_e),
\ee
which is the standard Schwarschild-AdS mass. 
However, numerical integration of the
field equations (\ref{bhhairy14d})-(\ref{bhhairy24d}) shows 
that for a large class
of boundary conditions there is in addition a one-parameter family of static
spherically symmetric black hole solutions with scalar hair outside the 
horizon \cite{Hertog04}.

The value $\phi_e$ of the field at the horizon as a function of horizon size
$R_e$ is plotted in Figure 1. The two curves correspond to solutions with two 
different AdS invariant boundary conditions, namely $f=-1$ (bottom) and 
$f=-1/4$ (top). Generically, we obtain $\phi_e>0$ if $f<0$ and $\phi_e<0$ 
for $f>0$. Only for $f=0$ and $f \rightarrow \infty$ there exist
no regular hairy black hole solutions.

%%%%%%%%%%%%%%%%%%%%%%%%%%%%%%%%
\begin{figure}[htb]
\begin{picture}(0,0)
\put(34,125){$M_h/4\pi$}
\put(204,14){$R_e$}
\put(245,148){$M_h/M_s$}
\put(455,12){$R_e$}
\end{picture}
\mbox{
\epsfxsize=7cm
\epsffile{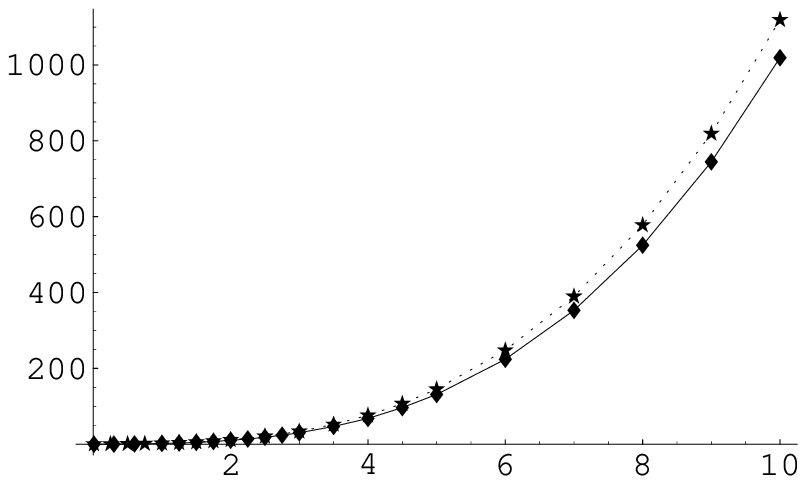}
\raisebox{2.3cm}{~~~~
\begin{minipage}{10cm}
\epsffile{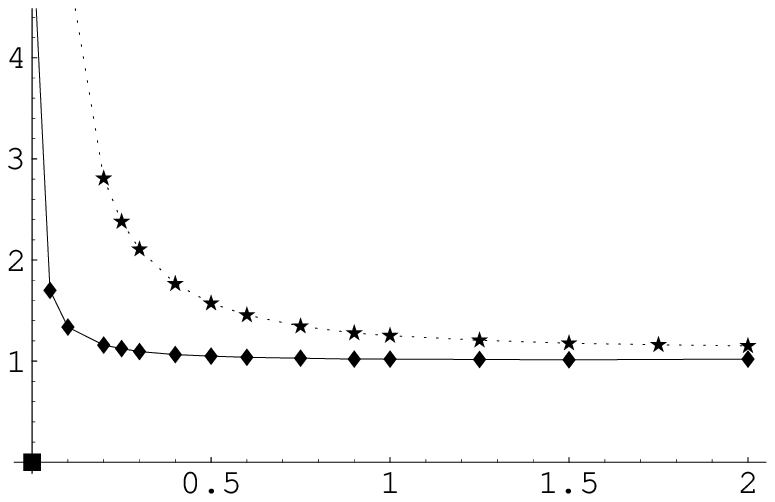}
\end{minipage}}}
\caption{left: The total mass $M_{h}/4\pi$ of hairy black holes as a 
function of horizon size $R_e$, in $D=4$ ${\cal N}=8$ supergravity with
two different AdS invariant boundary conditions $f=-1/4$ (top) and $f=-1$ 
(bottom). 
right: The ratio $M_{h}/M_{s}$ as a function of horizon size $R_e$, where 
$M_s$ is the mass of a Schwarschild-AdS black hole of the same size $R_e$}
\label{2}
\end{figure}
%%%%%%%%%%%%%%%%%%%%%%%%%%%%%%%%%%%
The integration constant $M_0$ in (\ref{asmetric4d})
is proportional to the finite gravitational contribution 
to the mass. It is, however, of little physical significance. Indeed the total
gravitational mass diverges. The relevant quantity is the conserved charge 
$Q[\partial_{t}]$, which is given by
\be \label{mass4dhair}
M_{h} =Q[\partial_{t}]=4\pi \left( M_0+\frac{4}{3}f\alpha^3 \right).  
\ee
The total mass $M_h$ is shown in Figure 2 as a function of horizon 
size $R_e$ and for two different boundary conditions $f=-1/4$ (top) and 
$f=-1$ (bottom). We find $M_h >0$ for all $R_e$ and for all boundary conditions
we have considered. For large $R_e$ one has $M_h \sim R_e^3$.
The mass is also compared with the mass $M_{s}$ of a Schwarschild-AdS black 
hole of the same size $R_e$. We find $M_{h}/M_{s} >1$ for all $R_e$ and 
$M_{h}/M_{s} \rightarrow 1$ for large $R_e$.

For given AdS invariant boundary conditions, there is at most one 
hairy black hole solution for a given total mass $Q[\partial_{t}]$,
so the horizon size as well as the value of the scalar field at the horizon
are uniquely determined by $Q[\partial_{t}]$. Thus we have found a 
one-parameter family of black holes with scalar hair, in a class of theories
parameterized by $f$. Because Schwarschild-AdS is a solution too for all 
boundary conditions we have two very different black hole 
solutions for a given total mass, one with $\phi=0$ everywhere and 
one with nontrivial hair. The scalar no hair theorem, therefore, 
does not in general hold in $D=4$ ${\cal N}=8$ supergravity with 
asymptotically 
anti-de Sitter boundary conditions. Uniqueness is restored only in theories 
with $f=0$ or for $f \rightarrow \infty$.
The stability and thermodynamic properties of these hairy black hole 
solutions is currently under investigation \cite{Hertog04d}.

\setcounter{equation}{0}
%%%%%%%%%%%%%%%%%%%%%%
\section{Solitons}
%%%%%%%%%%%%%%%%%%%%%

The existence of hairy black holes suggests there should also be regular 
static, spherically symmetric solitons that obey the same boundary
conditions (\ref{4-scalar})-(\ref{4-grr}). Soliton solutions can 
similarly be 
found by numerically solving eqs (\ref{bhhairy14d}-\ref{bhhairy34d}). 
Regularity at 
the origin now requires $h=1$, $h_{,r}=0$ and $\phi_{,r}=0$ at $r=0$.

%%%%%%%%%%%%%%%%%%%%%%%%%%%%%%%%
\begin{figure}[htb]
\begin{picture}(0,0)
\put(34,258){$\phi$}
\put(440,21){$r$}
\end{picture}
\mbox{\epsfxsize=15cm \epsffile{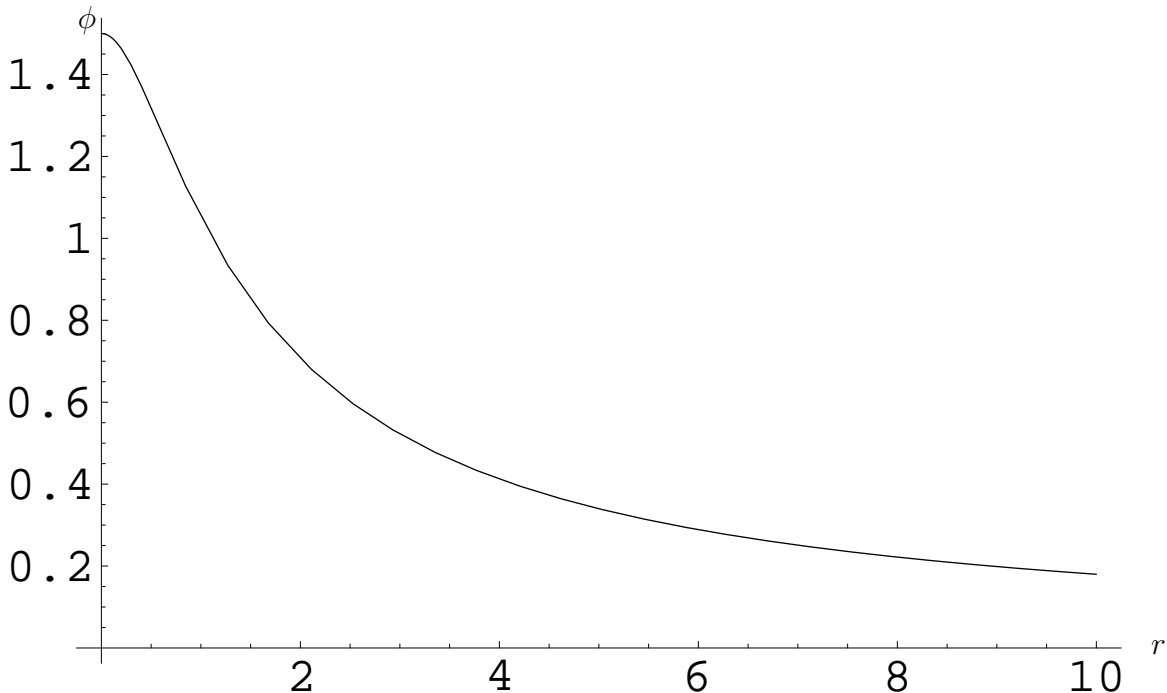}}
\caption{Soliton solution $\phi(r)$ in $D=4$ supergravity
with boundary conditions specified by $f=-1/4$.}
\label{3}
\end{figure}
%%%%%%%%%%%%%%%%%%%%%%%%%%%%%%%%%%%

For every nonzero $\phi_0$ at the origin, the solution
to (\ref{bhhairy14d}) is asymptotically of the form (\ref{4-scalar}) 
for some value of $f$. The staticity and spherical symmetry of the soliton 
mean $\alpha (t,x^a)$ is simply a constant.
The scalar field value $\phi_0$ at the origin uniquely 
determines $f$ and vice versa: there is at most one static spherical soliton 
solution in each theory. We find \cite{Hertog04c} there is
a regular soliton solution for all finite 
$f \neq 0$. When $|f| \rightarrow 0$ one finds
$ |\phi_0|  \rightarrow \infty$ and
for $|f| \rightarrow  \infty$ one has $ |\phi_0|  \rightarrow 0$ so the
nontrivial soliton solution ceases to exist in this limit.
As an example, in Figure 3 we show the soliton solution for
$f=-1/4$ boundary conditions, which has $\phi_0 \approx 1.5$. 

Most importantly, the existence of soliton solutions for a large class of 
AdS invariant boundary conditions implies 
supergravity with these boundary conditions does not admit a positive mass
theorem \cite{Hertog04c}. This can be seen as follows. For the spherical
solitons the constraint equation (\ref{bhhairy24d}) can be integrated, which
yields a formal expression for the gravitational surface term 
(\ref{gravcharge})
\be\label{gensoln} 
Q_G [\partial_{t}] =  2\pi
\lim_{r\to\infty}\int_{0}^{r} e^{-{1\over 2}
\int_{\tilde r}^r  d\hat r \ \hat r\pr^2} 
\left[2(V(\phi)-\Lambda) +\left(1+ {\tilde r^2 \over \ell^2}\right)
\phi_{,\tilde r}^2 \right] \tilde r^2 d\tilde r.  
\ee
One must add to this the scalar surface term to obtain the mass
(\ref{charge4d}). Now consider a family of configurations 
$\phi_{\lambda}(r) = \phi_0 (\lambda r)$ with mass $M_{\lambda}$
where $\phi_0 (r)$ is the static soliton profile.
From (\ref{gensoln}) and the form of the scalar contribution 
one sees that the soliton mass $M$ consists of the sum of a finite term 
$M_1$ (which includes the scalar contribution) that 
scales as the volume under rescalings $r \rightarrow \lambda r$ and a finite
term $M_2$ that scales linearly in $r$. The latter comes from
the gradient terms $\phi'^2_{,r}$ in (\ref{gensoln}) and is manifestly 
positive. Therefore, one has
\be
M_\l = \l^{-3} M_1 + \l^{-1} M_2
\ee
Since the soliton extremizes the mass it follows that
\be
\frac{dM_{\lambda}}{d\lambda}=-3M_1 -M_2=0
\ee
Hence $M_1$ must be negative for the soliton. But this means rescaled 
configurations $\phi_\l(r) = \phi_0(\l r)$ with sufficiently small $\l$  
have negative mass. The AdS solution is unstable, therefore, with 
generalized boundary conditions (\ref{4-scalar}) on the negative 
$m^2$ scalar.

Usually one discards unstable theories, saying they are not of physical
interest. But here there should be a field theory dual to these bulk 
theories even if they are unstable. By studying the dual field 
theory description of various manifestations of the instability in 
the bulk, one can hope to gain insight into the quantum nature of 
such phenomena. Supergravity with generalized AdS invariant boundary 
conditions together with AdS/CFT thus provides a controlled setting to 
explore string theory away from the supersymmetric moduli space, where the 
theory is stable. In the next sections we further explore this instability,
concentrating on applications to cosmology. Finally
in section 8 we turn to the dual field theory description of this theory.

\setcounter{equation}{0}
%%%%%%%%%%%%%%%%%%%%%%%%%%%%
\section{Instantons}
%%%%%%%%%%%%%%%%%%%%%%%%%%%%

The existence of negative mass solutions means there must also be nontrivial
zero mass solutions. The best known examples of such solutions are obtained
from Euclidean instanton solutions which are usually interpreted as 
describing the decay of a false vacuum.
An $O(4)$-invariant instanton solution takes the form
\be \label{inst}
ds^2 = {d\r^2\over b^2(\r)} +\r^2 d\Omega_3
\ee
and $\phi=\phi(\r)$. The field equations determine $b$ in terms of $\phi$
\be \label{inst1}
b^2(\r) = { 2V \r^2 -6\over \r^2 \phi'^2 -6}
\ee
and the scalar field $\phi$ itself obeys
\be\label{inst2}
b^2 \phi'' + \left( {3 b^2 \over \r} +bb' \right) \phi' -V_{,\phi} =0
\ee
where prime denotes $\partial_{\rho}$. Regularity again requires $\phi'(0) =0$.

%%%%%%%%%%%%%%%%%%%%%%%%%%%%%%%%
\begin{figure}[htb]
\begin{picture}(0,0)
\put(37,263){$\phi$}
\put(440,21){$\r$}
\end{picture}
\mbox{\epsfxsize=15cm \epsffile{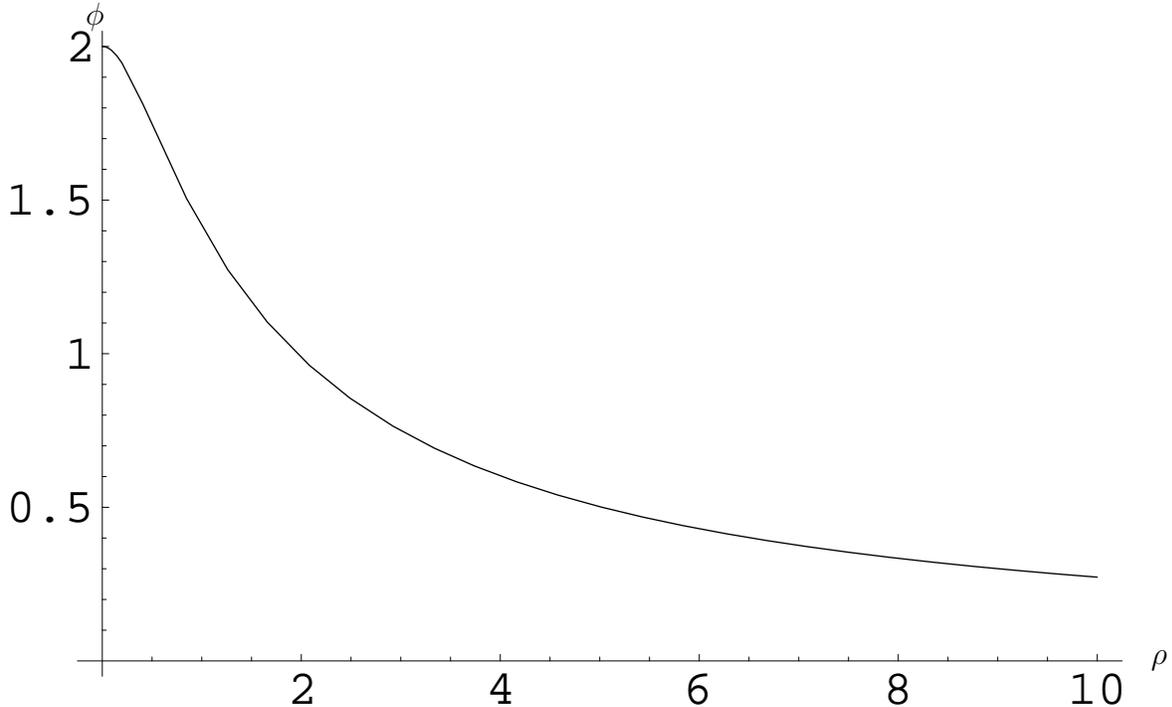}}
\caption{Instanton solution $\phi(\r)$ in $D=4$ supergravity
with boundary conditions specified by $f=-1/4$.}
\label{4}
\end{figure}
%%%%%%%%%%%%%%%%%%%%%%%%%%%%%%%%%%%

From (\ref{inst2}) it follows that asymptotically $\phi(\r)$ has the same
behavior as the Lorentzian scalar field solutions considered above,
\be \label{as-scalar}
\phi = {\alpha\over \r} + {f \alpha^2\over \r^2}.
\ee
We find that all boundary conditions that admit a spherical soliton solution
also admit an $O(4)$-invariant instanton solution. As for the solitons, 
 $f$ is determined by the field $\phi(0)$ at the origin. 
In Figure 4, the profile $\phi(\r)$ is shown of the instanton 
with $f=-1/4$ boundary conditions.

The instanton also defines a Lorentzian solution which is obtained by 
analytical continuation
across the equator of the three sphere. The fields on this slice of the 
instanton define time symmetric initial data for a Lorentzian solution. The 
Euclidean radial distance $\r$ simply becomes the radial distance $r$ in the 
Lorentzian solution. The total mass (\ref{charge4d})
of this initial data can be computed from the instanton geometry.
Substituting (\ref{as-scalar}) into (\ref{inst1}) yields asymptotically
\be\label{as-b}
b^2(\r) = \r^2  +1 +{\alpha^2 \over 2} + {4f\alpha^3\over 3 \r}
\ee
This is of the form (\ref{4-grr}) required to have 
finite conserved charges. In fact, we see that $M_0= -4f\alpha^3/3$ and hence
(\ref{charge4d}) implies that the total mass is zero! This is
consistent with the interpretation of the instanton as the solution 
$AdS_4$ decays into.

The quantum decay rate is determined in a semiclassical approximation by the 
Euclidean action of instanton. The action is given by
\be
I = \int [-{1\over 2} R + {1\over 2}(\nabla \phi)^2 + V(\phi)]
  - \oint K +{1 \over 6} \oint \left[ (\nabla \phi )^2 -m^2\phi^2 \right]    
\ee
where the first surface term is the usual Gibbons-Hawking term, and the
second is the surface term required so that the Hamiltonian constructed
from this action (after subtracting the background) agrees with 
(\ref{charge4d}).

The relevant quantity for computing the rate of vacuum decay is the
difference between the instanton action and the action for pure AdS:
$\Delta I = I - I_{AdS}$. Subtracting $I_{AdS}$ removes the leading
divergences in $I$, but since $\phi$  goes to zero so slowly,
there are two subleading divergences. If the coefficients of these terms
were not exactly zero, $\Delta I$ would be infinite and there would
be no probability for the vacuum to decay. We have shown \cite{Hertog04c} 
that both coefficients miraculously vanish. 
This involves nontrivial cancellations among the
volume term and both surface terms in the action. 
Furthermore, the difference $\Delta I$ becomes small for large $\vert f \vert$
and goes to zero when $\vert f \vert \rightarrow \infty$.

\setcounter{equation}{0}
%%%%%%%%%%%%%%%%%%%%%%%%
\section{Big Crunch Instability}
%%%%%%%%%%%%%%%%%%%%%%%

We now turn to the evolution of the state AdS decays into. This is in light 
of the AdS/CFT correspondence potentially
the most interesting manifestation of the supergravity instability. 
We will show that with generalized AdS invariant boundary conditions,
there are supergravity solutions where regular initial data evolve to 
a big crunch singularity. 

First let us return to the class of configurations $\phi_\l(r) = \phi_0(\l r)$,
where $\phi_0$ is the soliton profile discussed in section 5. 
The rescaled configurations $\phi_\l(r)$ specify initial data for 
time-dependent solutions in the same theory (i.e. with the same value of $f$).
For large $\l$, the initial bubble is smaller than the soliton and probably 
collapses. On the other hand, by taking $\l$ small one can arrange to have 
initially an arbitrarily large central region where $\phi$ is essentially 
constant and away from the maximum of the potential. It follows that the 
field must evolve to a spacelike singularity \cite{Hertog03a}. 
But the singularity that develops cannot be hidden behind an event horizon, 
because all spherically symmetric black holes have positive mass\footnote{We 
have demonstrated this for the $f=-1/4$ and $f=-1$ theories in section 4,
but this is true in general.} \cite{Hertog04} while the total mass of the 
rescaled initial data is negative.
Hence there is simply not enough mass to form a black hole, which encloses the
singular region. Instead, one expects the singularity to continue to spread, 
cutting off all space\footnote{If $V$ were bounded from below, it has been 
shown that the singularity cannot end or become timelike 
\cite{Dafermos:2004ws}. The same is likely to be true here.}. 
Boundary conditions that admit a soliton solution, therefore, also 
admit solutions where finite mass configurations produce a big crunch.

A particular example of such a solution where the evolution
is known explicitly is provided by the Euclidean instanton.
The evolution of initial data defined by slicing the instanton
across the three sphere is simply obtained by analytic
continuation. This is discussed in detail in
\cite{Coleman:1980aw}, but the basic idea is the following.
The origin of the Euclidean instanton
becomes the lightcone of the Lorentzian solution. Outside the
lightcone, the solution is given by (\ref{inst}) with $d\Omega_3$ replaced by
three dimensional de Sitter space. The scalar field $\phi$ remains bounded in
this region. Inside the lightcone, the $SO(3,1)$ symmetry ensures that
the solution evolves like an open FRW universe,
\be \label{ametric}
ds^2 = -dt^2+ a^2(t) d\sigma_3
\ee
where $d\sigma_3$ is the metric on the three dimensional unit hyperboloid.
The field equations are
\be\label{fieldeq}
{\ddot a\over a}= {1\over 3}[V(\phi) - \dot\phi^2]
\ee
\be\label{phieq}
\ddot{\phi} +{3 \dot a\over a} \dot\phi +V_{,\phi} =0 
\ee
and the constraint equation is
\be\label{constraint}
\dot a^2 - {a^2\over 3}\left [{1\over 2} \dot \phi^2 + V(\phi)\right ] = 1 \ ,
\ee
where $\dot a =\partial_{t} a$.
On the light cone, $\phi=\phi(0)$ and $\dot \phi=0$ (since $\phi_{,\r}=0$
at the origin in the instanton). 
Under evolution $\phi$ rolls down the negative potential, so the right
hand side of (\ref{fieldeq}) decreases. This ensures that $a(t)$ vanishes
in finite time
producing a big crunch singularity. For the purpose of understanding
cosmological singularities in string theory, one can forget the
origin of this solution as the analytic continuation of an instanton.
We have simply found an explicit example of asymptotically AdS initial
data which evolves to a big crunch.

We close this section with some comments on possible generalizations.
In section 2 we have shown that one can generalize the boundary conditions
while preserving the asymptotic AdS symmetries whenever one has a scalar 
field with $m^2_{BF}\le m^2<m^2_{BF}+1$ which decouples from the rest of the 
matter. In particular, this includes $\N=8$ supergravity 
in five dimensions, which involves a scalar field saturating the BF bound. 
In all cases one can construct similar solutions where a big crunch is 
produced from smooth finite mass initial data \cite{Hertog04c}.
The simplest solutions of this kind that we have presented here are 
constructed from time symmetric initial data, so they have a big
bang singularity in the past as well.
It would be interesting to construct solutions with only one singularity, 
in the future or the past. 

\setcounter{equation}{0}
%%%%%%%%%%%%%%%%%%%%%%
\section{Dual CFT description}
%%%%%%%%%%%%%%%%%%%%%%%

Having shown that the bulk theory admits solutions which evolve to
a big crunch, we now turn to the dual CFT description of this theory.
The dual to string theory on $AdS_4\times S^7$ can be obtained by
starting with the field theory on a stack of $N$ D2-branes. This is a
$SU(N)$ gauge theory with seven adjoint scalars $\vp^i$. One then takes
the infrared (strongly coupled) limit to obtain the CFT. In the process,
one obtains an $SO(8)$ symmetry. In the abelian case, $N=1$, this can be
understood by dualizing the three dimensional gauge field to obtain another
scalar. But in general, it is not well understood. 

This theory has dimension one operators $\O_T=Tr T_{ij} \vp^i\vp^j$ where
$T_{ij}$ is symmetric and traceless \cite{Aharony:1998rm}.\footnote{Since there are only seven
$\vp$'s and the theory has $SO(8)$ symmetry, there are other operators
involving the gauge field which complete the $SO(8)$ representation.}
One of these, $\O$, is dual to the bulk field we
have been considering with the boundary conditions that
$\phi = \alpha/r +O(r^{-3})$ for physical states. The field theory
dual to the ``standard" quantization, where physical states are described by
modes with $\phi =\beta/r^2$ asymptotically, can be obtained by adding
the double trace term ${f\over 2}\int \O^2$ to the action
\cite{Witten02,Gubser:2002vv}.
This is a relevant perturbation and the infrared limit is another CFT
in which $\O$ has dimension two. 

The AdS invariant boundary conditions we have considered here
correspond instead to adding a triple trace term to the action
\be\label{Ocubed}
S = S_0 + {f\over 3} \int \O^3
\ee
This follows from Witten's treatment of multi-trace operators in AdS/CFT
\cite{Witten02}. 
The extra term in  (\ref{Ocubed}) has dimension three, and hence is marginal 
and preserves conformal invariance, at least to leading order. One might 
wonder if this symmetry is exact, or whether
the operator $\O^3$ has an anomalous dimension. The anomalous dimension
can receive contributions proportional to $1/N$ or $f$. Since the large $N$
limit corresponds to supergravity in the bulk with AdS invariant boundary
conditions, and for every $f$ there is a bulk solution corresponding to 
pure AdS, it seems likely that the theory remains conformally invariant
for finite $f$ (at least for large $N$).

More generally, Witten's procedure says that all AdS invariant boundary 
conditions discussed in Section 2 are dual to field theories that differ from 
each other by multi-trace deformations preserving conformal invariance. Thus 
one obtains a line of conformal fixed points in each case\footnote{In 
supergravity theories with more than one scalar with 
$m^2_{BF}\le m^2<m^2_{BF}+1$ the different lines of conformal fixed 
points are parameterized by several dimensionless constants $f_i$.}.

We now turn to the dual field theory evolution of the big crunch solutions
considered above.
The Lorentzian solution obtained from the instanton takes the form
(\ref{inst}) with $d\Omega_3$ replaced by
three dimensional de Sitter space, $dS_3$. So
one might think that the natural dual would
correspond to the CFT on  $dS_3$. This
field theory certainly allows evolution for infinite time and is
nonsingular. But this only corresponds to evolution for finite global time.
We want to conformally rescale
$dS_3$ to (part of) the cylinder $R\times S^2$. This is equivalent to
a coordinate transformation in the bulk. The relation between the usual
static coordinates (\ref{adsmetric}) for $AdS_4$ and the $SO(3,1)$ invariant 
coordinates
\be
ds^2 = {d\r^2\over 1+\r^2} + \r^2 (-d \tau^2 + \cosh^2\tau d\Omega_2)
\ee
is 
\be
\r^2 = r^2 \cos^2 t -\sin^2 t
\ee
Since our bulk solution asymptotically has
\be
\phi(\r) = {\alpha\over \r} + {f\alpha^2\over \r^2}+ O(\r^{-3})
\ee
This becomes
\be
\phi(r) = {\tilde \alpha\over r} + {f\tilde\alpha^2\over r^2} +O(r^{-3})
\ee
where $\tilde\alpha= \alpha/\cos t$. Notice that $f$ is unchanged.
Hence the evolution of the initial data defined by the instanton
preserves the AdS invariant boundary conditions (\ref{4-scalar})-(\ref{4-grr}).
The fact that $\tilde\alpha$ blows
up as $t\rightarrow \pi/2$ is consistent with the fact that this is the
time that the big crunch singularity hits the boundary. The 
coefficient of $1/r$ is usually interpreted as the expectation value of $\O$
in the CFT. Hence AdS/CFT predicts that in the large $N$ approximation
the latter diverges too.

A qualitative explanation for this is the following.
The term we have added to the action is not positive definite. Since the
energy associated with the asymptotic time translation in the bulk can
be negative, the dual field should also admit negative energy states. This 
strongly suggests that the usual vacuum is unstable. It might
decay via the (nongravitational) decay of the false vacuum. 
Perhaps a useful analogy is
a scalar field theory with potential $V= m^2 \vp^2 - f\vp^6$. The quadratic
term is analogous to the  coupling of $\vp$ to the curvature of $S^2$, which
is needed for conformal invariance. The second term is analogous to the
second term in (\ref{Ocubed}). Qualitatively
this theory has the same behavior as the bulk. There
are instantons which describe the semiclassical decay of the usual
vacuum at $\vp=0$. For small $f$, the potential barrier is large, and the
instanton action is large. So tunneling is suppressed. For large $f$,
the barrier is small and tunneling is not suppressed. Classically one finds 
that after the tunneling the field rolls down the potential and 
becomes infinite in finite time. This means that in the semiclassical 
description of this analogous field theory, evolution ends in finite time. 
The fact that the field becomes infinite in this scalar field theory is
analogous to the divergence of the expectation value of $\O$ in the theory
(\ref{Ocubed}). Whether this means that evolution ends in the full quantum 
description of the CFT remains a fascinating open question, which we are 
currently investigating. If so, one could conclude that there is no bounce 
through the big crunch singularity in the bulk.

\setcounter{equation}{0}
%%%%%%%%%%%%%%%%%%%%%%
\section{Conclusion}
%%%%%%%%%%%%%%%%%%%%%%

We have studied solutions of $\N=8$, $D=4$ supergravity where the $m^2=-2$
scalar is the only excited matter field. Since its mass lies in the range 
$m^2_{BF} \le m^2 <m^2_{BF} +1$, 
there is a one-parameter family of boundary conditions on the scalar (and the 
metric components) that preserve the full AdS symmetry group. When the 
parameter vanishes, the dual CFT is the usual $2+1$ theory on a stack of 
M2-branes. Nonzero values of the parameter correspond to modifying this theory
by a triple trace operator. We find that for all nonzero values, there exists 
a family of AdS black holes with scalar hair. Both the horizon size
of the hairy black hole solutions and the value of the scalar at the 
horizon are uniquely determined by a single conserved charge, namely the mass.
Since Schwarschild-AdS is a solution too for all boundary 
conditions, one has two very different black hole solutions for a given
total mass. The uniqueness or no scalar hair theorem, therefore, does not 
hold in supergravity with generalized AdS invariant boundary conditions. 
It would be interesting to see how a microscopic string theory description
distinguishes between both classes of black hole solutions.

Although the modified boundary conditions preserve the full set of asymptotic 
AdS symmetries and allow a finite conserved energy to be defined, we have
shown this energy can be negative. Thus the AdS solution in supergravity
with these boundary conditions is nonlinearly unstable. A particular 
manifestation of this is that there are asymptotically AdS solutions 
describing the evolution of regular finite mass initial data to a big crunch. 

Our motivation to study supergravity in this regime is that there should be a 
dual CFT description of these bulk theories even if they are unstable.
Most interestingly, the field theory should provide a complete quantum 
description of the big crunch singularity. If states in the CFT have
a well defined evolution for all time, and one can reconstruct from it 
a semiclassical bulk metric at late time, then there must be a bounce through
the singularity in the full string theory. However, if the CFT evolution ends 
after finite time, or a semiclassical metric cannot be constructed, then the 
bulk evolution would end at the big crunch. 

As we mentioned, modifying the bulk boundary conditions corresponds to 
modifying the usual dual
field theory on a stack of M2-branes by a triple trace operator. Since this 
term is not positive definite, it appears possible there will be certain CFT 
states which do not have well defined evolution for all time. 
We have seen this happening at the semiclassical level in the deformed $2+1$ 
theory for states that are dual to our big crunch supergravity solutions. 

Moreover, we have good evidence that there are no bulk solutions that produce
a big crunch in supergravity theories that are dual to stable CFT's. This is 
because solutions of this type would violate\footnote{There is no naked 
singularity, but one does not have well defined evolution for all time in the 
asymptotic region.} cosmic censorship \cite{Penrose69}, which is believed to 
hold in theories with a positive mass theorem, even in anti-de Sitter space
\cite{Hertog03b,Hertog04b} (see also \cite{Gutperle04,Hubeny:2004cn}.
Taken together, these results indicate that producing a big crunch in AdS
from smooth initial data {\em requires} boundary conditions that correspond 
to an unstable dual CFT. Therefore, the fact that the dual classical evolution
ends in finite time and that the expectation value of the operator $\O$ dual
to the bulk scalar field
diverges in the large $N$ limit are, presumably, generic properties of a dual 
field theory description of a big crunch in the bulk, at least 
with AdS boundary conditions. Whether this means the big crunch is 
an endpoint of evolution in the full string theory remains a fascinating
open question, which we are currently investigating.
If it is an endpoint, that would raise 
the issue what determines the boundary conditions at 
cosmological singularities. Perhaps the AdS/CFT correspondence and the 
toy models of cosmologies we have constructed here could be useful to study 
this question further.

\bigskip

\centerline{{\bf Acknowledgments}}

Special thanks to my collaborators G. Horowitz and K. Maeda for their 
assistance on the work presented here. I also wish to thank C. Pope and H. Lu 
of the Mitchell Institute for Fundamental Physics at Texas A\&M University for
their hospitality and for organizing a stimulating conference. 
This work was supported in part by NSF grant PHY-0244764.

\end{document}